\documentclass[useAMS,usenatbib]{mn2e}
\usepackage{graphicx}
\usepackage{amssymb}
\makeatletter
\usepackage{times}
\usepackage{amsmath}

\newcommand{\peryr}{\,{\rm yr^{-1}}}

\newcommand{\yr}{{\,\rm yr}}
\newcommand{\cm}{{\,\rm cm}}

\newcommand{\Myr}{{\,\rm Myr}}

\newcommand{\perMyr}{{\,\rm Myr^{-1}}}
\newcommand{\pc}{\,\mathrm{pc}}
\newcommand{\kpc}{\,\mathrm{kpc}}
\newcommand{\Gpc}{\,\mathrm{Gpc}}
\newcommand{\Hz}{{\,\rm Hz}}
\newcommand{\mHz}{{\,\rm mHz}}

\newcommand{\Mbh}{M_{\bullet}}
\newcommand{\Ro}{R_{\odot}}
\newcommand{\Mo}{M_{\odot}}

\newcommand{\Rs}{R_{\star}}
\newcommand{\Ms}{M_{\star}}

\newcommand{\LISA}{{\it LISA\,}}

\newcommand{\apj}{ApJ}
\newcommand{\apjl}{ApJL}

\newcommand{\prd}{Phys. Rev. D.}
\newcommand{\mnras}{MNRAS}
\newcommand{\aap}{A.A.P.}
\newcommand{\physrep}{Phys. Rep.}
\newcommand{\nat}{Nature}

\title[Gravitational wave bursts from the Galactic massive black
hole]{Gravitational wave bursts from the Galactic massive black hole}

\author[Clovis Hopman, Marc Freitag and Shane L. Larson]
{
Clovis Hopman$^1$\thanks{e-mail: clovis@strw.leidenuniv.nl}, Marc
Freitag$^2$ and Shane L. Larson$^3$\\
$^1$Leiden University, Leiden Observatory, P.O. Box 9513, NL-2300 RA
Leiden\\
$^2$  Institute of Astronomy, University of Cambridge, Madingley
Road, CB3 0HA Cambridge, UK\\
$^3$Department of Physics, Weber State University, Ogden, UT   84408
}

\begin{document}
\bibliographystyle{mn2e.bst}

\date{Accepted ????. Received ?????; in original form ?????}

\pagerange{\pageref{firstpage}--\pageref{lastpage}} \pubyear{2006}

\maketitle

\label{firstpage}

\begin{abstract}
The Galactic massive black hole (MBH), with a mass of
$\Mbh=3.6\times10^6\Mo$, is the closest known MBH, at a distance of
only $8\kpc$. The proximity of this MBH makes it possible to observe
gravitational waves from stars with periapse in the observational
frequency window of the {\it Laser Interferometer Space Antenna}
(\LISA). This is possible even if the orbit of the star is very
eccentric, so that the orbital frequency is many orders of magnitude
below the \LISA frequency window, as suggested by \citet{Rub06}. Here
we give an analytical estimate of the detection rate of such
gravitational wave bursts. The burst rate is critically sensitive to
the inner cut-off of the stellar density profile. Our model accounts
for mass-segregation and for the physics determining the inner radius
of the cusp, such as stellar collisions, energy dissipation by
gravitational wave emission, and consequences of the finite number of
stars. We find that stellar black holes have a burst rate of the order
of {{}}$1\peryr$, while the rate is of order {{}}$\lesssim0.1\peryr$
for main sequence stars and white dwarfs. These analytical estimates
are supported by a series of Monte Carlo samplings of the expected
distribution of stars around the Galactic MBH, which yield the full
probability distribution for the rates. We estimate that no burst will
be observable from the Virgo cluster. 
\end{abstract}


\section{Introduction}

When a star comes near the event horizon of a massive black hole (MBH)
with mass $\Mbh\lesssim{\rm few}\times10^6\Mo$, it emits gravitational
waves (GWs) with frequencies observable by the planned {\it Laser
Interferometer Space Antenna} (\LISA).  Such extreme mass ratio
inspiral sources (EMRIs) can be observed by \LISA to cosmological
distances, provided that they spend their entire orbit emitting GWs in
the \LISA frequency band \citep{Bar04a, Bar04b, Gai04, Fin00, Gla05}.
For an EMRI to be in the \LISA band, the orbital period of the star
has to be shorter than $P\sim10^4{\rm s}$.  The formation mechanism
for EMRIs begins when a star that is initially not strongly bound to
the MBH is scattered to a highly eccentric orbit, such that its
periapse comes close to the Schwarzschild radius $r_S$ of the MBH. The
star loses energy to GWs on every orbit, slowly spiraling inward.
This process may eventually lead to a closely bound orbit that is
observable by \LISA. Inspiral is often frustrated by scattering with
other field stars \citep{Ale03b, Hop05}, and the rate at which stars
manage to spiral in successfully is rather low, of the order of
$0.1\perMyr$ per Galaxy for stellar black holes (BHs) (Hils \& Bender
\citeyear{Hil95}; Sigurdsson \& Rees \citeyear{Sig97b};
Miralda-Escud\'e \& Gould \citeyear{Mir00}; Freitag \citeyear{Fre01};
Ivanov \citeyear{Iva02}; Freitag \citeyear{Fre03}; Hopman \& Alexander
\citeyear{Hop06, Hop06b}; see Hopman \citeyear{Hop06c} for a review).
However, due to the large distances ($\sim1\Gpc$) to which EMRI
sources can be observed, the integrated rate over the volume makes
these a very promising target for \LISA.

Our own Galactic centre contains a MBH of
$\Mbh=(3.6\pm0.3)\times10^6\Mo$ \citep{Sch02, Eis05, Ale05}, in the
range of MBH masses that will be probed by \LISA. Since the Galactic
MBH is very close, $d\approx 8\kpc$ \citep{Eis05}, the stars near the
MBH can be studied in great detail \citep{Gen03a, Sch02, Sch03, Ghe03a,
Ghe05}. The Galactic MBH and its stellar cluster are therefore very
useful as a prototype for extra-galactic nuclei, in particular in the
study of EMRI sources. For a review of stellar processes near MBHs,
see \citet{Ale05}.

It is unclear whether our own Galactic centre (GC) can be observed as a
continuous source of GWs. From the very low event rates this appears
to be highly unlikely, but due to its proximity, waves with lower
frequencies can be observed in the GC, and it was suggested by
\citet{Fre03} that a number of low mass main sequence (MS) stars may
be observed in our own Galactic centre.

Another possibility was considered by Rubbo, Holley-Bockelmann \& Finn
(\citeyear{Rub06}; hereafter RHBF06), who showed that even a single
fly-by of a star near the MBH would be observable by \LISA if it is
sufficiently close, and they estimated that the event rate is high
enough ($\sim15\peryr$) that several detectable fly-bys would be
observable during the \LISA mission.  The prospect of observing the
Galactic MBH as a GW burster is very exciting: it would imply
detection of GWs from an object that has been extensively studied in
many electromagnetic wavelengths.  Furthermore, as we point out in
this paper, GW bursts are caused by stars very close to the MBH, and
thus probe a region near the MBH which is not accessible
observationally in a direct way by other means.

RHBF06 used a single mass stellar model to study the GW burst
rate. The rate is dominated by nearby stars, raising the question what
determines the inner cut-off of the stellar cusp; RHBF06 assumed that
the cusp extends all the way to the MBH. Here we re-address the event
rate of such GW bursts in the Galactic centre. We consider the
treatment of a multi-mass system, and account for the inner cut-off of
the cusp.

This paper is organized as follows.  In \S\ref{s:det} we derive the
minimal periapse a star needs to have to give an observable burst of
GWs in the Galactic centre.  In \S \ref{s:rate} we give an analytical
expression for the event rate of GW bursts.  The rate is dominated by
stars at very close distances from the MBH, and we discuss several
processes which may determine the inner cut-off of the density
profile.  Our analytical model is complemented by Monte Carlo
realisations of stellar cusps (\S\ref{s:MC}), which allow an accurate
treatment of rare events where a single star produces a large number
of bursts. The Monte Carlo samplings also yield the probability
distribution of the event rates, in addition to the average event
rate. In \S{\ref{s:results} we present our resulting rates, and in
\S\ref{s:disc} we discuss and summarize our results.

\section{Detection of gravitational waves from the Galactic
centre}\label{s:det}

For $f<\mHz$, a good approximation of the sensitivity
curve of \LISA \citep{Lar01} is given by

\begin{equation}\label{e:Sapprox}
S(f) =S_0\left({f\over {\rm Hz}}\right)^{-4}\left({L\over
5\times10^{11}\cm}\right)^{-1/2}\quad (f<\mHz)
\end{equation}
where $S_0 = 6.16\times10^{-51}\,\Hz^{-1}$, $f$ is the frequency of a
GW, and $L$ is the arm length of \LISA \citep{LHH00}.

The signal to noise ratio $\rho$ can be approximated by (Finn \&
Thorne \citeyear{Fin00}, Eq. [2.2])

\begin{equation}\label{e:SNR}
\rho \approx {h\mathcal{N}_{f}^{1/2}\over f^{1/2} S^{1/2}};
\end{equation}
here is $\mathcal{N}_{f}$ the number of cycles spent at a certain
frequency $f$; for GW bursts $\mathcal{N}_{f}=1$. Furthermore, $h$ is
the strain, which is here approximated by the quadrupole estimate of a
circular orbit (Finn \& Thorne \citeyear{Fin00}, Eq. [3.13])

\begin{eqnarray}\label{e:h}
h &=& 2.28\times10^{-16} \left({d\over 8 \kpc}\right)^{-1}\nonumber\\
&&\times \left({m\over\Mo}\right)\left({\Mbh\over
3.6\times10^6\Mo}\right)^{2/3}\left({f\over {\rm Hz}}\right)^{2/3},
\end{eqnarray}
From Eqs (\ref{e:Sapprox}, \ref{e:SNR}, \ref{e:h}), the minimal
frequency necessary to measure a burst is then

\begin{eqnarray}\label{e:fburst}
f_{\rm burst} &=& 4.3\times10^{-5}\,\Hz\,\rho^{6/13}\left({d\over 8
\kpc}\right)^{6/13}\nonumber\\
&&\times \left({m\over\Mo}\right)^{-6/13}\left({\Mbh\over
3.6\times10^6\Mo}\right)^{-4/13},
\end{eqnarray}
where $f=\sqrt{GM/r_p^3}$ is now the orbital frequency at periapse of
an eccentric orbit. The corresponding periapse is

\begin{eqnarray}\label{e:rpburst}
\left({r_p^{\rm burst}\over r_S}\right) &=&
60\,\rho^{-4/13}\left({d\over 8 \kpc}\right)^{-4/13}\nonumber\\
&&\times \left({m\over\Mo}\right)^{4/13}\left({\Mbh\over
3.6\times10^6\Mo}\right)^{-6/13},
\end{eqnarray}
or {{}}$r_p^{\rm burst}\approx2.1\times10^{-5}\pc$. The corresponding
angular momentum is

\begin{eqnarray}
\left({J_{\rm burst}\over J_{\rm LSO}}\right)^2 &=&
7.5\,\rho^{-4/13}\left({d\over 8
\kpc}\right)^{-4/13}\left({m\over\Mo}\right)^{4/13}\nonumber\\
&&\times \left({\Mbh\over
3.6\times10^6\Mo}\right)^{-6/13}\left(2-{r_p^{\rm burst}\over
a}\right)\,,
\end{eqnarray}
where $J_{\rm LSO}^2=(4G\Mbh/c)^2$ defines the last stable orbit.

The event rate (Eq.  \ref{e:gam}) is approximately proportional to
$J_{\rm burst}^2$.  In this model, we note that neglecting the noise
from Galactic white dwarf (WD) binaries can be justified by the fact
that the noise is much smaller than the instrumental noise at $f_{\rm
burst} = 3\times10^{-5}\Hz$ \citep{Hil97}.  At higher frequencies, $2
\times 10^{-4} {\rm Hz} \lesssim f \lesssim 3 \times 10^{-3} {\rm Hz}$
the SNR will increase in spite of the presence of galactic noise,
because of the larger GW amplitude at those frequencies.

\section{An analytical model for the gravitational wave burst rate in
the Galactic centre}\label{s:rate}

\subsection{The gravitational wave burst rate in an isotropic
distribution}

The distribution of stars near a MBH is an important problem in
stellar dynamics, and has been studied since the early seventies
\citep{Pee72}. The MBH dominates the dynamics within the radius of
influence, $r_h=G\Mbh/\sigma^2$, where $\sigma$ is the stellar
velocity dispersion far away from the MBH. For the Galactic centre,
$r_h\approx2\pc$ \citep{Hop06}. It was shown by \citet{Bah76} that
within $r_h$, the density distribution of a single mass population of
stars is very well approximated by a power-law, $n(r)\propto
r^{-\alpha}$, with $\alpha=7/4$. These results, which were obtained by
solving the Fokker-Planck equation in energy space, were later
confirmed by $N$-body simulations \citep{Bau04a, Pre04} and Monte
Carlo simulations \citep{Fre02}.

\citet{Bah77} studied the distribution of stars with different masses
near a MBH, and showed that mass segregation leads to steeper
distributions of the more massive species which sink to the centre due
to dynamical friction. These results were confirmed and extended by
\citet{Fre06} and \citet{Hop06b} for a much wider range of masses. For
simplicity we approximate the distributions as power-laws, with
different exponents for different species (symbolised by
``$M$''), such that $n_{M}(r)\propto
r^{-\alpha_{M}}$. The values of $\alpha_{M}$ will be discussed in
\S\ref{s:GC}.

We assume an isotropic density profile; the role of modifications of the
DF by the loss-cone will be discussed in \S\ref{s:lc}. For such a
distribution, the number of stars $n(a, dJ^2)dadJ^2$ in an element
$(a, a+da), (J^2, J^2+dJ^2)$ is given by

\begin{equation}\label{e:naJ2}
n(a, J^2)dadJ^2 = (3-\alpha_{M}){C_{M}N_h\over r_h}\left({a\over
r_h}\right)^{2-\alpha_{M}}{1\over J_c^2(a)}dadJ^2,
\end{equation}
where $J_c(a)=\sqrt{G\Mbh a}$ is the circular angular momentum, $N_h$
is the number of MS stars within $r_h$ and $C_{M}N_h$ the total number
of stars of type $M$ within $r_h$ (so that for MS stars $C_{\rm
MS}\equiv1$). The rate per unit of logarithmic of the semi-major
axis at which stars of species $M$ have a bursting interaction with
the MBH is then given by

\begin{equation}\label{e:gam}
a{d\Gamma_{M}\over da} = (3-\alpha_{M}){C_{M} N_h\over
P(a)}\left({a\over r_h}\right)^{3-\alpha_{M}}\left[{J_{\rm
burst}^2(a) - J_{\rm LSO}^2\over J_{c}^2(a)}\right],
\end{equation}
where $P(a)=2\pi(a^3/G\Mbh)^{1/2}$ is the period. For MS stars, the
term $J_{\rm LSO}^2$ in Eq. (\ref{e:gam}) should be replaced with the
tidal loss-cone, $J_t^2=2G\Mbh r_t$. Here $r_t=(\Mbh/\Ms)^{1/3}\Rs$ is
the tidal radius, where a star is disrupted by the tidal force of
the MBH.

Equation (\ref{e:gam}) gives an analytical estimate of the GW burst
rate in our Galactic centre, assuming that the distribution function
can be approximated as being an isotropic power-law distribution, with
different powers for different species.  It is only valid within the
radius of influence $r_h$.  Since the event rate is entirely dominated
by stars very close to the MBH (see Fig.  [\ref{f:rate}]), we neglect
contributions to the GW burst rate from stars with $a>r_h$.  From
equation (\ref{e:gam}) it can be seen that for the relevant values of
$\alpha>1/2$, the GW burst rate formally diverges for nearby stars
(RHBF06): setting $J_{\rm LSO}\to0$, the rate is proportional to
$a{\rm d}\Gamma_{M}/ {\rm d}a\propto a^{1/2 - \alpha}$.

Finally, we note that RHBF06 make a number of cuts in phase space;
these cuts are either made here implicitly, or they do not affect our
results.

\subsection{The stellar distribution in the Galactic
centre}\label{s:GC}

The stellar cluster in the Galactic centre has been observed in the
infra-red in much detail. It has been shown \citep{Ale99a, Gen03a,
Ale05} that the stars in the Galactic centre are distributed in a cusp
with profile $\rho\propto r^{-1.4}$ consistent with the predictions by
\citet{Bah76, Bah77}, although it is important to note that only the
most luminous stars can be observed, and that the observations are
therefore strongly biased. The stellar population at $1-100\pc$ is
consistent with a model of continuous star formation \citep{Ser96,
Fig04}. Within the radius of influence $r_h=2\pc$, \citet{Gen03a}
finds that there is a total mass $M_{\rm tot}=1.7\times10^6\Mo$ in
stars.

Not much is known observationally about the inner $\sim0.01\pc$ of the
Galactic centre. There are a number of B-stars (known as the ``S
cluster'') at that distance, which provide a challenge for star
formation theories, but it is not known whether these stars are
representative for the dimmer stars present there: it is more likely
that they are the result of tidal binary disruptions by the MBH
\citep{Gou03, Per06}. The S-stars can also be used to probe the
enclosed dark mass. This is how the total mass of the MBH,
$\Mbh=3.6\times10^6\Mo$ \citep{Eis05} can be determined, but in
principle the orbits of the S-stars can be used to constrain the
nature of the extended mass by looking for deviations of Keplerian
motion: if, for example, a cluster of stellar black holes is present,
the orbits of the S-stars should precess. To date, searches for
deviations from Keplerian motion do not lead to relevant constraints
\citep{Mou05}.

By lack of direct observations of the stellar content of the inner
region of the Galaxy, we resort to theoretical models for
mass-segregation. Such models were recently made by \citet{Fre06} and
\citet{Hop06b}, and show that stellar black holes have a much steeper
cusp than the other species. 

We consider 4 distinct species of stars: MS stars, WDs, neutron stars
(NSs) and stellar BHs, with $M_{\rm MS}=0.5\Mo$, $M_{\rm WD}=0.6\Mo$,
$M_{\rm NS}=1.4\Mo$, $M_{\rm BH}=10\Mo$. We assume that the enclosed
number of MS stars within the radius of influence at $r_h=2\pc$ is
$N_h=3.4\times10^6$, and that the number of compact remnants are equal
to that resulting from Fokker-Planck calculations by \citet{Hop06},
who found $C_{\rm MS}=1$, $C_{\rm WD}=0.14$, $C_{\rm
NS}=9\times10^{-3}$, $C_{\rm BH}=6\times10^{-3}$. For the slopes,
\citet{Hop06b} found $\alpha_{\rm MS}=1.4$, $\alpha_{\rm WD}=1.4$,
$\alpha_{\rm NS}=1.5$, $\alpha_{\rm BH}=2$. These slopes are all quite
different from those assumed by RHBF06, who assumed $\alpha=1.75$ for
all species.

\subsection{The inner region of the stellar cusp}\label{s:rin}

The rate of GW bursts is dominated by stars very close to the MBH. It
is therefore important to estimate to which distance the cusp
continues. Here we consider a number of processes that can determine
the inner edge of the cusp.

\subsubsection{Finite number effects}  

Current models of stellar systems near MBHs rely mainly on statistical
approaches such as Fokker-Planck methods; $N$-body simulations can
only be performed for small systems with intermediate mass black holes
(IMBHs) of masses $\Mbh\sim10^3\Mo$ \citep{Bau04a, Bau04b, Pre04}. In
particular, the \citet{Bah76, Bah77} solutions which first predicted
the slope of the stellar cusp, can in principle extend to any inner
radius if there is no physical mechanism that provides a cut-off (such
as stellar collisions, or tidal disruption). In reality, there is only
a finite number of stars; this implies that even if no inner cut-off
of the cusp is provided by a physical mechanism that destroys the
stars, there is an inner radius beyond which no stars are
expected. Statistically, the cusp runs out at\footnote{This expression
is also given in \citet{Hop06}, but with an error in the sign of the
exponent.}  

\begin{equation}\label{e:rstat}
r_{1, M}=(C_{M}N_h)^{-1/(3-\alpha_{M})}r_h. 
\end{equation}

Using $r_h=2\pc$, this gives for the favored model {{}}$r_{\rm 1,
MS}=2\times10^{-4}\pc$, $r_{\rm 1, WD}=6\times10^{-4}\pc$, $r_{\rm 1,
NS} = 2\times10^{-3}\pc$ and $r_{\rm 1, BH}=1\times10^{-4}\pc$.

We neglect in our analytical estimate contributions from rare cases
where there is a star within $r_{1,M}$. However, we do explore this
possibility in the Monte Carlo samplings presented in \S\ref{s:MC}.

\subsubsection{Hydrodynamical collisions}  

Close to a MBH, the number density and velocity dispersions become
very large, and stars will collide within their life-times
\citep{Fra76, Coh78, Mur91}. The rate $\Gamma_{\rm coll}$ at which
stars with radius $\Rs$ at a distance $r_h$ from the MBH have grazing
collisions can be estimated as

\begin{equation}\label{e:coll}
\Gamma_{\rm coll} = nv\Sigma = {3-\alpha\over 4\pi}{N_h\over
r_h^3}\left({r\over r_h}\right)^{-\alpha}\left({G\Mbh\over
r}\right)^{1/2}\pi\Rs^2,
\end{equation}
where $\Sigma=\pi\Rs^2$ is the cross-section for a grazing collision,
when the relative velocity is significantly larger than the escape
velocity from the surface of the star.

Studies by \citep{Fre02, Fre05} show a single grazing collision is
unlikely to disrupt a star, but that rather $N_{\rm coll}\sim20-30$
collisions are required for disruption \citep{Fre06}. This implies
that stars are disrupted by collisions within a Hubble time if
their distance from the MBH is smaller than

\begin{equation}\label{e:rcoll}
r_{\rm coll} = 3\times10^{-3}\pc\left({N_{\rm
coll}\over30}\right)^{1/2}\left({\Rs\over\Ro}\right),
\end{equation}
where it was assumed that $\alpha=3/2$, as is approximately the case
for MS stars.

Although this estimate is clearly not very precise, it is unlikely
that the cusp for MS stars continues much closer to the MBH than
$r_{\rm coll}$, since collisions become very frequent and with higher
impact velocities. For the preferred model we assume $r_{\rm
coll}$ as the inner cut-off of the cusp for MS stars. We do not
consider collisions between other stellar species.

\subsubsection{Gravitational wave inspiral}\label{s:insp} 

GW emission plays a double role: on the one hand the GWs can be
detected by \LISA, but on the other they also change the dynamics
close to the MBH, since the star emitting the GW loses orbital energy,
and spirals in. Close to the MBH, stars spiral in faster than they are
replenished by other stars. This region of phase-space is therefore
typically empty, because any bursting star would be quickly accreted
by the MBH.

If $r_p \ll a$, the inspiral time is approximately given by
\citep{Pet64} 

\begin{equation}\label{e:t0}
t_0(r_p, a)={2\pi\sqrt{G\Mbh a}\over \Delta E_{\rm GW}(r_p, a)},
\end{equation}
where
\begin{equation}
 \Delta E_{\rm GW}(r_p, a)={2\pi\over 5\sqrt{2}}f(e){\Ms
c^2\over\Mbh}\left({r_p\over r_S}\right)^{-7/2},
\end{equation}
and $f(e)\approx2.2$. If this time-scale at $r_p=r_p^{\rm burst}$ is
much smaller than the time-scale $t_J(r_p,a)\sim(r_p/a)t_r$ for
two-body scattering to change the angular momentum by order unity,
stars with $r_p<r_p^{\rm burst}$ spiral in much faster than they
are replenished. Solving $t_J(r_p^{\rm burst},a)=t_0(r_p^{\rm burst},
a)$ for $a$ gives an inner cut-off

\begin{equation}\label{e:a_insp}
{{}}a_{\rm GW} =
1.9\times10^{-4}\pc\,\left({m\over\Mo}\right)^{2/13}\left({d\over 8
\kpc}\right)^{20/39}\left({\rho\over 5}\right)^{20/39},
\end{equation}
where a relaxation time of $t_r=10^9\yr$ was assumed.

\subsubsection{Kicks out of the cusp} 
An important assumption that is routinely made in stellar dynamics is
that the rate at which stars exchange energy and angular momentum is
dominated by small angle encounters \citep[e.g.][]{Cha43, Bin87}.
This is justified by comparing the large-angle scattering timescale,
$t_{\rm LA}\approx [nv(G\Ms/v^2)^2]^{-1}$, to the relaxation time
$t_r$.  The timescale for large-angle scattering by a single strong
encounter is larger than the relaxation time (where many small
encounters add up to a large angle) by a factor $t_{\rm
LA}/t_r\sim\ln\Lambda$, where $\ln\Lambda$ is the Coulomb logarithm;
close to a MBH, $\Lambda\sim\Mbh/\Ms$ \citep{Bah76}.

In spite of this, large angle scattering may play an important role in
the ejection of stars out of the cusp \citep{Lin80, Bau04a}. Whether
the rate of ejections out of the cusp is larger than the rate at which
stars are swallowed by the MBH may depend on the size of the system:
\citet{Lin80} and \citet{Bau04a} find that the ejection rate is larger
for intermediate mass black holes of $\Mbh\sim10^3\Mo$, but the
swallow rate is much higher for MBHs \citep{Fre06}.

Even if the ejection rate is larger than the merger rate, in all cases
the rate at which stars are replenished by diffusion in energy space
is larger than the ejection rate by a factor $\ln\Lambda$
\citep{Bah77, Lin80}. Ejections can therefore never deplete the inner
region of the cusp, and need not be considered for the purposes of
this paper. We note that \citet{Bau04a} found that all stellar BHs are
ejected, but this cannot happen in a galactic nucleus where these
objects are constantly replenished by mass segregation from larger
radii.

\subsubsection{The role of the loss-cone}\label{s:lc}

We assume an isotropic velocity distribution for the stars, leading to
the DF $n(a,J^2)$ given in Eq. (\ref{e:naJ2}). We do not consider
stars in the region $J<J_{\rm LSO}$, which is the ``loss-cone'';
loss-cone theory shows that so close to the MBH, there are no stars in
this region in phase space, because any star will be immediately
removed \citep{Lig77, Coh78}.

 In reality, there will be a smooth transition from the empty region
of the loss-cone to the region far away from the loss-cone (large
angular momenta). \citet{Lig77} find that close to the loss-cone,
there is a logarithmic depletion of stars. Taking this factor into
account leads to a suppression of the GW burst rates by a factor of
order $\sim3$ compared to the results we present here. On the other
hand, resonant relaxation \citep{Rau96} may replenish some stars to
this region \citep{Rau98}, although the effect will be not very large
due to general relativistic precession, which destroys the resonant
relaxation.

In this paper we do not consider any modification of the DF by the
presence of the loss-cone, but note that this approach may be somewhat
optimistic.

\subsection{Main model}\label{s:mainmod}

To summarize, the method to compute the GW burst rate is as
follows. Four species of stars (MS, WD, NS, BH) are considered, all
with their own mass $m/\Mo=(0.5, 0.6, 1.4, 10)$, number normalization
$C_{M}=(1,0.14,0.009,6\times10^{-3})$ at $r_h=2\pc$, slope
$\alpha_{M}$, inner radius $r_{{\rm in}, M}$ and loss-cone
$J_{lc}=\max(J_{\rm LSO}, J_t)$. The total number of MS stars
($C_{M}=1$) within the cusp is $N_h=3.4\times10^6$. These values are
used in Eq. (\ref{e:gam}); the rate $\Gamma_{M}$ is then integrated to
find the total GW burst rate for each species.

For the model which is regarded to reflect the stellar population in
the Galactic centre best, the following values are assumed. For the
slopes, $\alpha_{M}=(1.4, 1.4, 1.5,2.0)$; for the inner
radius, $r_{\rm in}=\max(r_{\rm coll}, r_1, a_{\rm GW})${{}}, we found
$r_{\rm in, MS}=3\times10^{-3}\pc$ (collisions), $r_{\rm in,
WD}=6\times10^{-4}\pc$, $r_{\rm in, NS} =2\times10^{-3}\pc$ (finite
number effects), and $r_{\rm in, BH}=3\times10^{-4}\pc$ (GW
inspiral).

Some other models are also considered for the purpose of comparison.

\section{Monte Carlo realisations of stellar cusps}\label{s:MC}

The analytical method described in the previous section is useful to
obtain an estimate of the average burst rate. However, it discards
rare events where a single star comes very close to the MBH and has a
large number of bursts per year. It also does not give information on
the distribution of the burst rate. In order to obtain this
information we complemented our analytical estimate by a Monte Carlo
approach, in which we produce a large number of realisations of the
models discussed in the previous section. This gives us the cumulative
probability $P(>\Gamma)$ that the event rate is higher than
$\Gamma$. In the Monte Carlo samplings, we do not need to explicitly
include a cut-off at small radii to account for statistical
depletion. Stars with $(1-e)t_r>t_0(e,a)$ are discarded (see
eq. \ref{s:insp}). Other cuts are identical to those made in the
analytical approach.

One example of a realisation of the stellar cusp is shown in figure
(\ref{f:a-e}).

\section{Results}\label{s:results}

\begin{figure*}
  \resizebox{11.5cm}{!}{%
          \includegraphics{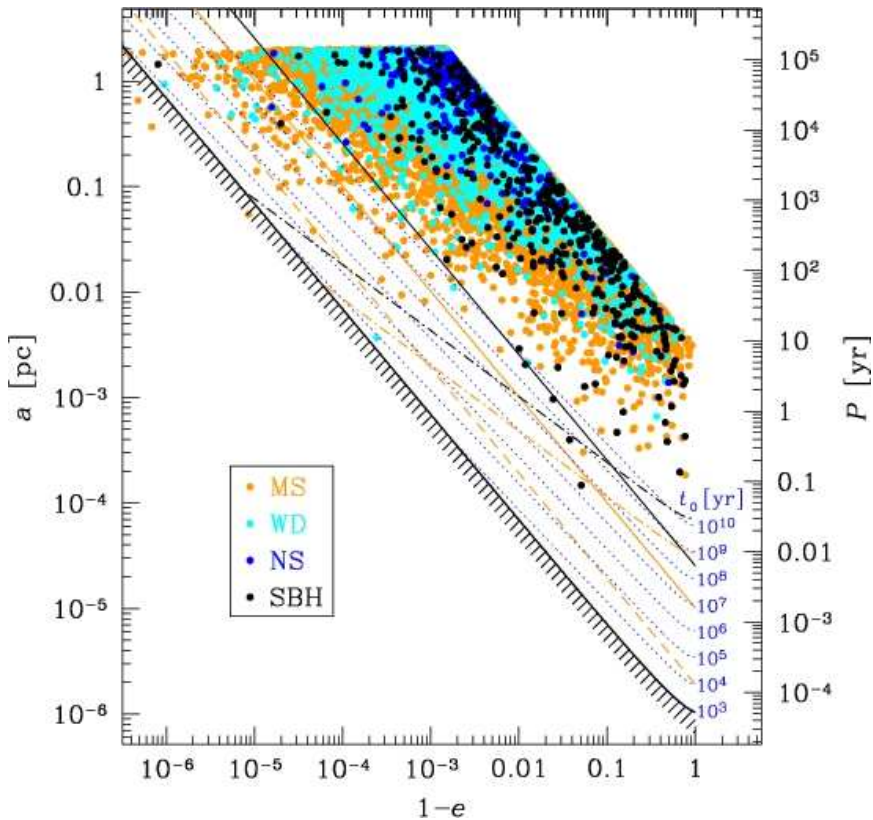}
        }
\hfill \parbox[b]{55mm}{%
\caption{Realisation of the main model for the stellar cusp, with
$1-e$ on the horizontal axis and the semi-major axis on the vertical
axis (or period on the right vertical axis). Orange dots represent MS
stars, cyan dots WDs, blue dots NSs, and black dots BHs, with numbers
and slopes as described in the main model. The black line with
hatching demarcates the last stable orbit. The dotted blue lines give
the GW inspiraling time (Eq. \ref{e:t0}) for the MS stars. The solid
orange line indicate the "burst region" for MS stars (with SNR=5). The
solid black line line above it is for stellar BHs. The dashed orange
line is the tidal disruption radius for MS stars. The orange
dot-dashed line shows $t_0=(1-e)t_r$ for MS stars while the black one
is for stellar BHs.  Below these lines, depletion by GW inspiral
should occur.  The region with $P<1$\,yr, where most bursting stars
should be is almost entirely depleted due to the finite number
effect.\label{f:a-e}} }
\end{figure*}

A number of different possibilities were considered for the slopes
and inner cut-offs of the respective stellar populations. The
resulting GW burst rates for these models are summarized in table
(\ref{t:t1}).

For our main model, we find that GW bursts are unlikely to be observed
in the Galactic centre for MS stars ({{}}$\Gamma_{\rm MS}\sim0.1\peryr$), for WDs
({{}}$\Gamma_{\rm WD}\sim0.1\peryr$) and for NSs ({{}}$\Gamma_{\rm
NS}\sim0.004\peryr$). Our burst rate for MSs is much lower than the
$\Gamma_{\rm MS}\sim12\peryr$ rate estimated by RHBF06; the main
reason for the difference is the cut-off due to collisions. Our rates
for WDs and NSs are also lower than those found by RHBF06 (who
estimated $\Gamma_{\rm WD}\sim3\peryr$ and $\Gamma_{\rm
NS}\sim0.1\peryr$); here the difference is probably caused mainly by
the different density profile, and the fact that the cusp runs out of
stars at small radii from the MBH. On the other hand, we find a higher
rate of BH bursts, ({{}}$\Gamma_{\rm BH}\sim1\peryr$), which is the
result of the steeper density profile we assumed, caused by
mass-segregation \citep{Fre06, Hop06b}. The inner radius for BHs was
determined by GW inspiral in this case (\S\ref{s:insp}).

The cumulative probability to detect more than a certain number of
bursts per year can be determined with Monte Carlo sampling (\S
\ref{s:MC}). It consists of several factors, including the probability
that in spite of the finite number effect a star has a very short
period in a certain realisation. In this latter case there is a large
number of correlated bursts, so that the distribution is not
Poissonian. We show the results for the main model in
Fig.~(\ref{f:P}). The average rates are in good agreement with our
analytical model. Smaller differences are that for WDs and NSs, the
Monte Carlo rates are slightly higher because of rare events excluded
in the analytical model, while for BHs the rates are slightly lower,
due to a small difference in the criterion for GW inspiral. From the
figure it can be confirmed that the probability to observe even one
single burst for MSs, WDs and NSs is negligible, but there is some
chance to observe several GW bursts from BHs.  The probability that
the rate of observed BH bursts per year is exceeds 1 is
{{}}$P(>1\,\peryr)\approx 20\%$. For illustration purposes, we show in
Fig~(\ref{f:a-e}) an example of a realisation of the main model.

\begin{figure}
\includegraphics[angle=0,scale=.4]{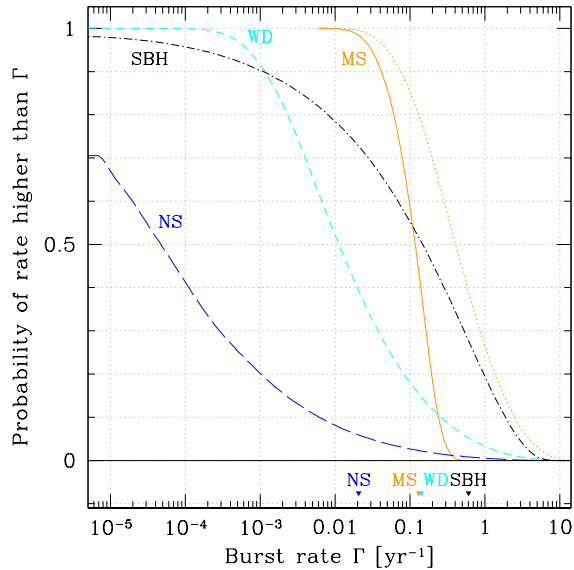}
\caption{Distribution of burst rates for 40\,000 Monte Carlo
realisations. We plot the probability for the burst rate of each
stellar species to be larger than a value $\Gamma$ as a function of
$\Gamma$. Each cusp realisation consists of $N_h=3.4\times 10^6$
particles distributed around an MBH according to the parameters of the
main model, such as illustrated in Fig.~(\ref{f:a-e}). Stars on plunge
orbits or with $(1-e)t_r > t_0(r_p,a)$ (see \S~\ref{s:insp}) are
discarded. We also remove MS stars with periapse distance smaller than
the tidal disruption radius $r_{t}\simeq 2\times 10^{-6}\,$pc or
semi-major axis smaller than the collision radius $r_{\rm
coll}=3\times 10^{-3}\,$pc. The dotted line indicate the rate
distribution for MS stars if there were no collisional depletion
($r_{\rm coll}=0$). The triangles above the horizontal axis indicate
the rates averaged over all realisations. Notice that, except for MS
stars, they are much higher than the median rates.}
\label{f:P}
\end{figure}

To probe the sensitivity of the GW burst rate to the assumptions, a
number of other possibilities are considered explicitly. \emph{We stress
that these models lack in realism; we consider them with the purpose
of probing how sensitive our results are to the assumptions made.}

First, consider the possibility that there is mass segregation, but an
inner cut-off of only $a_{\rm min}=3\times10^{-5}\pc$ for all stars
(this is approximately where bursting sources become continuous
sources, see eq. [\ref{e:rpburst}] and RHBF06). This increases the
rate considerably for all species. The GW burst rate is plotted in
Fig. (\ref{f:rate}). From this figure it can also be seen what the
event rates for the main model are, by considering the appropriate
cut-off for each species.

\begin{figure}
\includegraphics[angle=0,scale=.4]{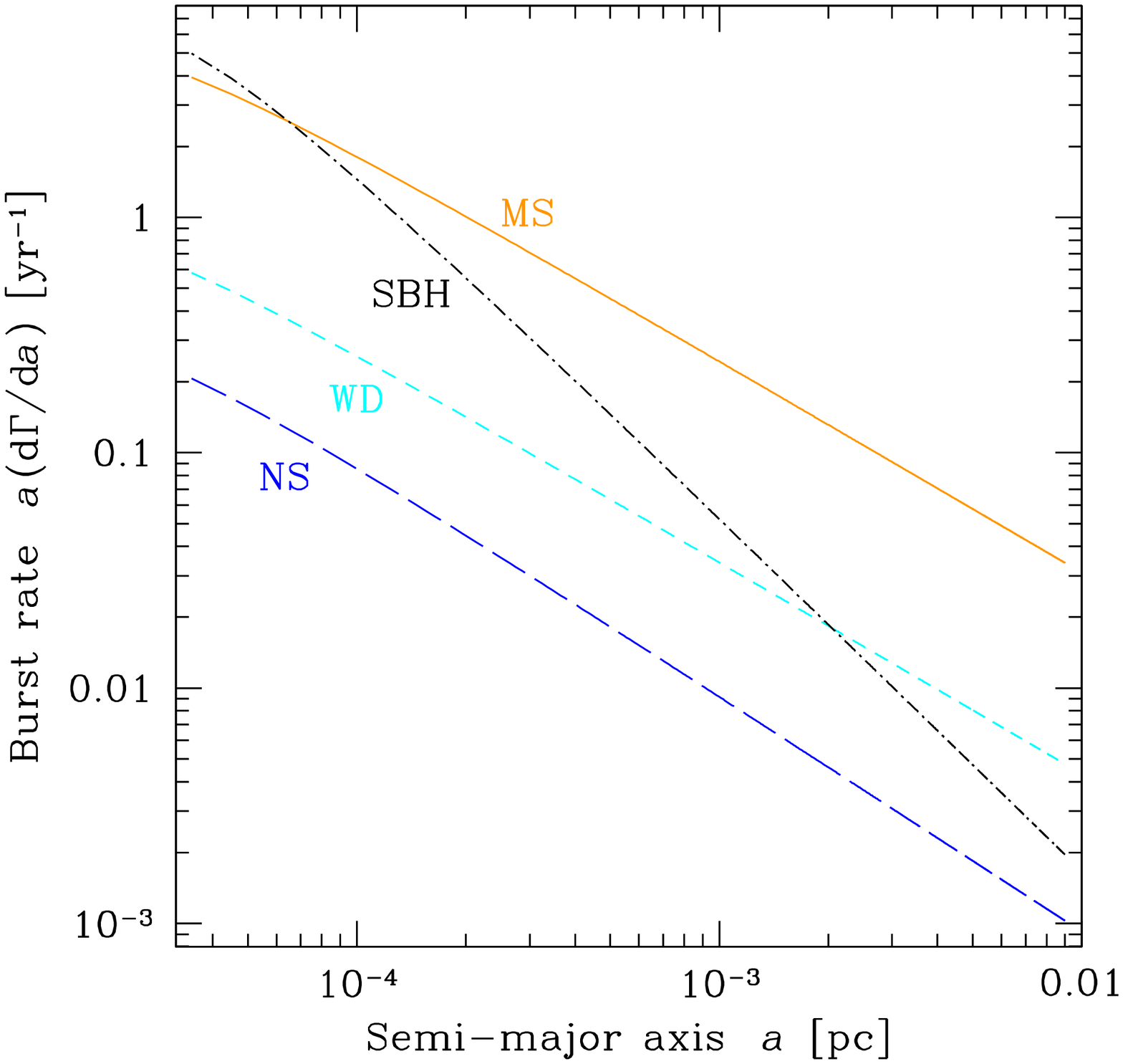}
\caption{{{}}The rate $a{\rm d}\Gamma_{M}/{\rm d}a$ for the preferred
model with $(\alpha_{\rm MS}, \alpha_{\rm WD}, \alpha_{\rm NS},
\alpha_{\rm BH}) = (1.4, 1.4, 1.5, 2)$ and $(C_{\rm MS}, C_{\rm WD},
C_{\rm NS}, C_{\rm BH}) = (1,0.14,0.009,6\times10^{-3})$.  The
required signal to noise ratio for detection was assumed to be
$\rho=5$.  The lines are extended for the case where there is no inner
cut-off to the distribution function.  In our main model, the rate
does have an inner cut-off, see table (\ref{t:t1}).  \label{f:rate}}
\end{figure}

Alternatively, we consider the case that there is an inner cut-off
equal for all species to that of the main model, but that the slope is
$\alpha_{M}=1.75$ for all species, as would be the case for a single
mass population. We used a normalization {{}}$(C_{\rm MS}:C_{\rm
WD}:C_{\rm NS}:C_{\rm BH})=(1:0.1:0.01:1\times10^{-3})$ here. This
example would present some realism if all stars are of similar mass
and in particular if the typical mass of stellar BHs would be of the
order $M_{\rm BH}\sim1\,\Mo$, or if the number of stellar BHs would be
much smaller than assumed here (in the latter case the burst rate of
the BHs would of course be lower). In this case, the GW burst rate
would be dominated by WDs, with a rate of the order of
{{}}$\Gamma_{\rm WD}\sim4\peryr$. Interestingly, the cut-offs for this
model are determined by different mechanism than for our main model
(see table \ref{t:t1}).

Finally, equation (\ref{e:gam}) was applied to model parameters
similar to those assumed by RHBF06, i.e., without mass-segregation and
with a fixed, very small inner cut-off. In this case an event rate is
found that is more than an order of magnitude above that found by
RHBF06. It is unclear what causes the discrepancy.

\begin{table*}
\begin{center}
\begin{tabular}{llllllllll}
\hline
\hline
Model                                    &  Star   &   $m$  & $C_M$
& $\alpha_{M}$ &  $a_{\rm in}$   & Reason for cut-off & $\Gamma_{M}$
\tabularnewline
                                         &         & $[\Mo]$&
&              &    $[\pc]$      &                    &   $[\peryr]$
\tabularnewline
\hline
Mass segregation, cut-off (main model)   & MS     &  0.5     & 1.0
&    1.4       & $3\times10^{-3}$ & Collisions         &   0.1
\tabularnewline
.......................................  & WD     &  0.6     & 0.14
&    1.4       & $6\times10^{-4}$ & Finite number      &   0.09
\tabularnewline
.......................................  & NS     &  1.4     & 0.009
&    1.5       & $2\times10^{-3}$ & Finite number      &
$4\times10^{-3}$  \tabularnewline
.......................................  & SBH    &  10      &
$6\times10^{-3}$  &    2.0       & $3\times10^{-4}$ & GW inspiral
&   1.5               \tabularnewline
\hline
Mass segregation, no cut-off             & MS     &  0.5     & 1.0
&    1.4       & $3\times10^{-5}$ & -                  &    6
\tabularnewline
.......................................  & WD     &  0.6     & 0.14
&    1.4       & $3\times10^{-5}$ & -                  &    0.8
\tabularnewline
.......................................  & NS     &  1.4     & 0.009
&    1.5       & $3\times10^{-5}$ & -                  &   0.3
\tabularnewline
.......................................  & SBH    &  10      &
$6\times10^{-3}$  &    2.0       & $3\times10^{-5}$ & -
&    5               \tabularnewline
\hline
No mass segregation, cut-off             & MS     &  0.5     & 1.0
&    1.75      & $3\times10^{-3}$ &  Collisions        &  0.5
\tabularnewline
.......................................  & WD     &  0.6     & 0.1
&    1.75      & $1\times10^{-4}$  &  GW inspiral      &   4
\tabularnewline
.......................................  & NS     &  1.4     & 0.01
&    1.75      & $5\times10^{-4}$  &  Finite number     &  0.1
\tabularnewline
.......................................  & SBH    &  10      &
$1\times10^{-3}$  &    1.75      & $3\times10^{-3}$ &  Finite number
&  $2\times10^{-3}$  \tabularnewline
\hline
No mass segregation, no cut-off          &MS     &  0.5     &
1.0          &    1.75      & $3\times10^{-5}$ &  -                 &
280                \tabularnewline
.......................................  & WD     &  0.6     & 0.1
&    1.75      & $3\times10^{-5}$  &  -                 & 42
\tabularnewline
.......................................  & NS     &  1.4     & 0.01
&    1.75      & $3\times10^{-5}$  &  -                 & 5
\tabularnewline
.......................................  & SBH    &  10      &
$1\times10^{-3}$  &    1.75      & $3\times10^{-5}$ &  -
& 0.7                \tabularnewline
\hline
\end{tabular}
\caption{Event rates for a number of different stellar species and
slopes.  For all cases the required signal to noise ratio for
detection was assumed to be $\rho=5$.  The first four entries give the
favored model, which accounts for mass-segregation according to the
results by \citet{Hop06b}, and has an inner cut-off of the cusp due to
stellar collisions or finite number effects.  The following four
entries give the same model, but with equal inner cut-offs $r_{\rm
in}=3\times10^{-5}\pc$ for all stars.  The next four entries are
without mass-segregation, but with an inner cut-off; this could be
appropriate if there are no SBHs (although they do appear in the
table), in which case mass-segregation would be much less extreme.
The last four entries also have the same inner cut-off, and an equal
slope $\alpha=1.75$ as appropriate for a single mass cusp, similar to
what was assumed by RHBF06.\label{t:t1}}
\end{center}
\end{table*}

\section{Summary and discussion}\label{s:disc}

When stars come very close ({{}}$r_p\lesssim60r_S$, see
Eq. \ref{e:rpburst}) to the MBH in our Galactic centre, they emit a
burst of GWs that could be observable by \LISA (RHBF06). In this paper
an analytical estimate for the burst rate is given. The estimate
includes physics that was not considered by RHBF06, in particular
mass-segregation and processes which determine the inner cut-off of
the stellar distribution function. Mass-segregation mostly leads to different
contributions from different species. However, since the event rate is
dominated by stars very near the MBH (Eq. \ref{e:gam}), the inner
cut-of leads to a strong suppression of the GW burst rate. We find
that only stellar BHs have a reasonable chance of being observed as
bursting sources, with a rate of the order of {{}}$\Gamma\sim1\peryr$
for signal to noise $\rho=5$.

The stellar distribution function in the inner $0.01\pc$ is not known
in the Galactic centre, and the results presented here rely on
theoretical estimates, rather than on observations.  The role of
collisions on the inner structure of the cusp is still poorly known,
and if the inner cut-off would be considerably smaller than assumed
here, the GW burst rate for MS stars grows substantially.  Observation
of a number of GW bursts from the Galactic centre would therefore have
implications for our understanding of stellar dynamics near MBHs.  The
observation of a GW burst would probably allow one to constrain the
masses of the system. In our models, we find that stellar BHs are the
most likely candidates to be bursting sources.  However, if the
bursting source is a WD, then this would imply that either stellar BHs
have masses much lower than $10\Mo$, or that their number is much
smaller than assumed here; in both cases the distribution of WDs would
be steeper than we assumed, and our model with cut-off, but without
mass-segregation, indicates that several WD bursts per year are then
to be expected. Interestingly, similar conclusions would apply for
inspiral sources \citep{Hop06b}.

Using stellar dynamics simulations, \citet{Fre03} suggested that, at
any given time there are $\sim 1-3$ {\em continuous} GW sources at the
Galactic centre (i.e., EMRIs), namely MS stars with a mass of $\sim
0.05-0.1\,M_\odot$ on orbits with $P\lesssim 3\times 10^4\,$s. This
result would imply a burst rate much higher than estimated here. We
note that a large population of low-mass MS stars would lead to a
slightly higher burst rate because of the larger number of stars and
the weaker depletion by GW emission. However, the EMRI rates obtained
by \citet{Fre03} seem to have been overestimated, due to the
approximate treatment of the condition for GW-driven inspiral relying
on a noisy particle-based estimate of the relaxation
time. Furthermore, in that work, once it had reached the GW-dominated
regime the possibility for a MS star to be destroyed by collisions was
neglected.

We stress that a star on route to become an EMRI is unlikely to be a
bursting source: the event rate at which EMRIs are created is of order
$\Gamma_{\rm EMRI}\sim0.1\perMyr$, while the time which a typical
future EMRI spends at orbits with periods less than $1\yr$ is
$t_i\sim0.05\Myr$ \citep{Hop06b}, implying that the probability of
observing such a source in the GC is of order $\sim t_i\Gamma_{\rm
EMRI}\sim5\times10^{-3}$. This confirms our assumption that the inner
regions of the cusp are depleted in presence of GW energy losses
(\S\ref{s:insp}).

Bursts of GWs from stars passing close to extra-galactic MBHs are a
potential source of noise for \LISA. An estimate of the contribution
to \LISA's noise budget is out of the scope of this paper, and will be
considered elsewhere.

RHBF06 also considered the possibility of observing GW bursts from the
Virgo cluster, and estimated that only stellar BHs could be observed
as bursting sources, with of the order of 3 bursts per year.  We note
that our higher rate of bursting BHs in the centre of our Galaxy than
that found by RHBF06 does {\it not} imply that we also predict a
higher rate of bursts from Virgo: for fixed signal to noise, a smaller
periapse is required in Virgo, which in turn implies a larger inner
cut-off of the density profile (see eq.  \ref{e:a_insp}).  Taking this
into account, we find that the bursting rate in Virgo is only of the
order of $\sim10^{-4}\peryr$ per galaxy, yielding a negligible rate
for the Virgo cluster.

\section*{Acknowledgments}
Discussions with Louis Rubbo, Kelly Holley-Bockelmann and Cole Miller
are highly appreciated. We also thank Pau Amaro-Seoane for organizing
the EMRI conference at the Albert Einstein Institute in Golm, which
allowed us to discuss the ideas of this paper. C.H. was supported by a
Veni scholarship from the Netherlands Organization for Scientific
Research (NWO).  The work of M.F. is funded through the PPARC rolling
grant at the Institute of Astronomy (IoA) in Cambridge.  S.L.L. and
M.F. acknowledge the hospitality of the Center for Gravitational Wave
Physics at Penn State during the early stages of this work.
S.L.L. acknowledges partial support from the Center for Gravitational
Wave Physics (supported by the National Science Foundation under
cooperative agreement PHY 01-14375), and from NASA award NNG05GF71G.

\bibliographystyle{mn2e}

\bsp
\label{lastpage}
\end{document}